\documentclass[12pt,a4wide]{article}
\usepackage{epsfig}
\usepackage{graphicx}
\usepackage{amsmath,amscd,amsfonts,eucal,latexsym,amssymb,mathrsfs,multicol}
 \newtheorem{Definition}{Definition}[section]
 \newtheorem{Theorem}[Definition]{Theorem}
 \newtheorem{Proposition}[Definition]{Proposition}

\newsymbol\rest 1316         

\newcommand{\dirop}{{\boldsymbol{\nabla}}\!\!\!\!\!/ \,}
\newcommand{\nslash}{n \!\!\!/ \,}

\newsymbol\bt 1202  

\newcommand{\hide}[1]{} 

\begin{document}
\noindent
\begin{center}
{ \large \bf Quantum Dirac Field on Moyal-Minkowski Spacetime --- Illustrating
 Quantum Field Theory over Lorentzian Spectral Geometry}
\\[20pt]
{\sc Rainer Verch}\\[10pt]
Institut f\"ur Theoretische Physik,\\
Universit\"at Leipzig,\\
04009 Leipzig, Germany\\
e-mail: verch@itp.uni-leipzig.de
\end{center}
${}$\\[26pt]
{\small {\bf Abstract} 
A sketch of an approach towards Lorentzian spectral geometry (based on
joint work with Mario Paschke) is described, together with a general way
to define abstractly the quantized Dirac field on such Lorentzian spectral
geometries. Moyal-Minkowski spacetime serves as an example. The
scattering of the quantized Dirac field by a non-commutative (Moyal-deformed)
action of an external scalar potential is investigated. It is shown that differentiating
the S-matrix with respect to the strength of the scattering potential gives rise
to quantum field operators depending on elements of the non-commutative
algebra entering the spectral geo\-metry description of Moyal-Minkowski spacetime,
in the spirit of  ``Bogoliubov's formula'', in analogy to the situation found in 
external potential scattering by a usual scalar potential.
}
\section{Introduction}
The reason why we entertain the idea that non-commutative (NC) geometry provides
a description of spacetime structure which supersedes the picture of spacetime
as a differentiable manifold resides in the expectation that,
at extremely short distances\,/\,high energies the classical concept of ``events''
looses its meaning.
Hence, the mathematical concept of ``events'' as points in a smooth manifold would
no longer be appropriate.
The general argument leading to this expectation roughly runs as follows.
According to general relativity, the energy content of matter determines spacetime geometry.
The energy content of matter at very high energies and short distances
is described by quantum field theory.
Thus, we should expect a ``quantum description'' of spacetime geometry 
(and eventually, a theory of 
quantum gravity). 
 A characteristic feature of such a description 
would be uncertainty relations for spacetime localization
of events (marking ``placement of energy/matter'') 
to avoid matter from undergoing gravitational collapse, which would preclude 
any information about matter distribution/\,geometry.
This idea was  made precise by Doplicher, Fredenhagen and Roberts
\cite{DFR}.

The corresponding uncertainty relations for localization of events can be implemented by
requiring commutation relations for their spacetime coordinates.
In consequence,
the commutative algebra of coordinate functions of a ``classical'' spacetime manifold
is replaced by a non-commutative algebra, generated by a set of ``non-commutative coordinates''
and their commutation relations.

Following this philosophy, one can think of quite a number of different ways 
to set up ``NC spacetime coordinates''. Here is a sample of the more prominent
approaches which have been proposed and investigated: 
\\[6pt]
{\bf Moyal-deformed Minkowski spacetime} (see \cite{Szabo} and refs.\ cited there)
$$ { [X^\mu ,X^\nu ] = i\lambda \theta^{\mu \nu} \quad \text{with} \quad \theta^{\mu\nu} \in 
\mathbb{R} }$$
\\[6pt]
{\bf Lie-algebra deformation of Minkowski-spacetime}
$$ { [X^\mu ,X^\nu ] = i \lambda \tau^{\mu \nu}{}_\varrho X^\varrho \,, \quad
\tau^{\mu\nu}{}_{\varrho} \in \mathbb{C}} $$
(generalization: \quad
 ${[X^\mu ,X^\nu ]}  {= } {F(X)} \,, $ where $F$ is a sufficiently nice function, e.g.\ with
 \begin{eqnarray*}
 { F(X)} & {=} & 
{i\lambda \theta^{\mu\nu} + i \lambda \tau^{\mu\nu}{}_{\varrho} X^{\varrho} +
   i\lambda^2\sigma^{\mu\nu}{}_{\varrho \kappa}X^{\varrho}X^{\kappa} + O(\lambda^3,(X)^3)}
   \\
   \text{and / or} & & {\theta^{\mu\nu} = \theta^{\mu\nu}(X)}\,, \ \text{etc \ )}  
   \end{eqnarray*}
${}$\\
{\bf Quantum space or Hopf-algebraic deformed Minkowski spacetime}
$$ { X^\mu X^\nu = \lambda\xi^{\mu\nu}{}_{\varrho \kappa}X^{\varrho}X^{\kappa}\,, \quad
\xi^{\mu\nu}_{\varrho \kappa} \in \mathbb{C}}$$
${}$\\
{\bf DFR-Minkowski spacetime} \cite{DFR}
$${ [X^\mu,X^\nu] =
 i \lambda Q^{\mu \nu}\,, \quad [X^\kappa,Q^{\mu\nu}] = 0 \,.}$$
${}$
In these relations, $\lambda$ is some real parameter setting the scale at which
non-commutativity is relevant (supposing the other quantities characterizing non-commutativity
are roughly of order 1).
Of course, this list is not meant to be complete.

For all these various ``models'' of NC spaces, certain quantum field theoretic models have 
been studied on these NC spaces.
The general observation drawn from those investigations is that the
UV behaviour improves, but there are new types of IR problems.
In some approaches, they can be cured. 
Some work points at possibility of constructing interacting quantum field theory (QFT) models 
e.g.\ on Moyal-deformed spaces \cite{HesSi,GW,RV-TW}.

While this surely opens some very promising perspectives, there are also some
drawbacks and conceptual problems:
\begin{itemize}
\item[$\star$] 
The promising constructive work uses {\it Euclidean} NC Moyal spacetime. 
For this class of spacetimes (and other NC spacetimes) there 
is no counterpart to the Osterwalder-Schrader theorem which 
establishes a correspondence between QFTs on Euclidean space and QFT on Minkowski space.
\item[$\star$]
For Moyal deformed Minkowski spacetime, Lorentz covariance is broken (to a smaller
covariance group). This is conceptually unsatisfactory (although regarded by some as a
sign that Lorentz covariance {\it is} broken in nature).
\item[$\star$]
The {\it operational significance} of NC spacetime in its relation to the QFT on it
is often not very clear.
\item[$\star$] What replaces the locality concept which 
is central to QFT in Minkowski spacetime on an NC spacetime?
\item[$\star$] There are (more or less) good arguments for all of the various models of NC
spaces (spacetimes). Which is the most appropriate (if any)? 
What conceptual and mathematical framework is needed to stage a 
systematic discussion of this question?
\item[$\star$] What about general covariance? General relativity is one of the main
motivations for considering NC spacetime. In QFT on classical spacetime, one can
formulate general covariance for QFTs. This requires to consider not just a few particular
spacetime models, but a whole class of spacetimes (abstractly characterized --- ``model independent'').
\item[$\star$] Actually, what is a QFT on an NC spacetime? What are its 
characterizing properties (needed for a sound physical interpretation)? Is there
a model-independent framework --- model-independent both on the NC geometry side {\it and}
on the QFT side?
\end{itemize}

\section{Lorentzian Spectral Geometry (but only \\ some daring first steps into
a vast jungle)}
In an attempt to find answers to the list of questions displayed above, one
may invoke a framework which unifies the general features of NC geometries
as a starting point. In fact, there 
is a model-independent approach to (compact) Riemannian NC geometry ---
the spectral geometry approach developed by Alain Connes 
\cite{Connesbigpaper,Connes2,Connes,GVF}.

The mainstream opinion, at least among connoiseurs of the spectral geometry
approach, is that most of the examples of NC spaces usually considered
 (when they correspond to NC
generalizations of Riemannian geometries) fulfill the conditions of spectral geometry.
As it stands, this statement isn't fully correct, e.g.\ Euclidean Moyal space
does not correspond to an NC compact spectral geometry (since Euclidean
space is non-compact), and the setting of Connes needs to be adapted to
this case (see, e.g., \cite{GGBISV}). Thus, this mainstream opinion is subject
to making adaptations to the original setting of Connes, and to be fair, I am
unaware of any systematic investigation that would substantially support the 
stated opinion (and clarifies which adaptations have to be made in detail in the
various cases).
However, we take it, for the time being, as working hypothesis. 

The strength of the spectral geometry approach is based on
\begin{itemize}
\item  ``Naturality'' of the axioms
\item Structural theorems, including ``reconstruction''
of a compact Riemannian manifold with spin structure in the ``classical case''
\end{itemize}
Up to now, it remains unclear if a spectral geometry approach of comparable strength
can be developed for the case of semi-Riemannian NC geometries. 
There are, however, some approaches 
\cite{Hawkins,Moretti,Strohmaier,KopfPaschke,PaschkeVerch}. We will sketch here the 
approach outlined in \cite{PaschkeVerch} (which draws partially on \cite{Strohmaier} and
\cite{KopfPaschke}) 
and set forth in \cite{BorVe};
it is developed with a view on ``general covariance'' as a
central principle for quantum field theories on (NC) manifolds (cf.\ \cite{BFV,PaschkeVerch}
for discussion, see also discussion below). That approach has largely been developed
by Mario Paschke, together with the present author, but as yet, it is tentative and 
unfinished. It should be seen as a proposal in which direction a generalization
of compact Riemannian spectral geometry could proceed. Some structural elements can
be generalized quite straightforwardly, others are less clear, and 
out of the the various possibilities of generalization one has to make choices. 

Let us briefly remind the reader of the spectral geometry setting generalizing compact Riemannian
spin manifolds. The central structure is called a {\it spectral triple}, since initially
the emphasis was on a collection of three objects, but nowdays it has become
customary to list in fact five objects, yet still referring to their collection as a spectral triple. 
This understood, a spectral triple consists of a collection $(\mathcal{A}, \mathcal{H},
D, \overset{\circ}{\gamma}, J)$ where $\mathcal{H}$ is a Hilbert space,
$\mathcal{A}$ is a unital $*$-algebra of bounded linear operators on $\mathcal{H}$,
$D$ is an unbounded selfadjoint operator on a suitable dense domain in $\mathcal{H}$,
$\overset{\circ}{\gamma}$ (often denoted simply as $\gamma$) is a bounded 
operator on $\mathcal{H}$ while $J$ (often denoted as $C$) is a conjugation on 
$\mathcal{H}$. These objects are interrelated by a list of relations and regularity
conditions (see \cite{Connesbigpaper}). It turns out that, in the case that
$\mathcal{A}$ is abelian, the spectral triple is equivalent to a compact Riemannian
spin manifold, where $\mathcal{A}$ corresponds to the algebra of scalar $C^\infty$
functions on the manifold, $\mathcal{H}$ is the Hilbert space of $L^2$ sections of
the spinor boundle, $D$ is a Dirac operator (the principal symbol is unique),
$\overset{\circ}{\gamma}$ corresponds to an orientation and $J$ is charge conjugation
on the spinors \cite{Connesbigpaper}.
(For an alternative approach, which isn't related to spin structure and Dirac operators,
see again \cite{Connesbigpaper}.)
 When trying to generalize the structure
to the Lorentzian case, there are a number of difficulties.  
First, causally well-behaved Lorentzian manifolds (i.e.\ spacetimes) are non-compact
in timelike directions, so as to avoid closed causal curves. This makes it 
necessary to work with some non-unital algebras in place of $\mathcal{A}$,
and this entails other difficulties. (Moreover, one needs to unitalize some of these
algebras in the end, and there is no unique way of doing this, so some choice is
involved here). Secondly, in the Riemannian case the Dirac operator
on a spin manifold $D$ is
elliptic, which is quite important in the spectral geometry setting, but this is clearly not
the case of a Lorentzian spin manifold. So, for the Lorentzian setting one needs a way
of gaining an elliptic operator out of the Dirac operator. Moreover, on a Lorentzian
spin manifold there is no canonical (or covariant) scalar product on the sections
of the spinor bundle and thus no natural $L^2$ Hilbert space structure. However,
there is a canonical sesquilinear form $< f,h>$ on the $C_0^\infty$ sections $f,h$ of the spinor
bundle, induced by the operation of taking the Dirac adjoint of a spinor, 
and the Dirac operator of the Lorentzian spin manifold, which we will
denote here by $\dirop$ (following physicists' notation), is symmetric with respect to
this sesquilinear form on $C^\infty_0$ sections. When one takes a future pointing
unit vector field, $n$, on the Lorentzian spin manifold, then 
\begin{align} \label{scalprod}
(f,h) &= \ \ < f,\boldsymbol{\gamma}(n)h >
\end{align}
 yields a scalar product (or a
negative definite inner product, depending on choice of metric signature) on the 
$C_0^\infty$ sections $f,h$ of the spinor bundle, where $\boldsymbol{\gamma}(n)$
denotes the Clifford action of $n$ on the spinors
\cite{DimockD,BaerGinoux}. (In physicists' abstract index
notation, $\boldsymbol{\gamma}(n) = n^a\boldsymbol{\gamma}_a{}^A{}_B$, or
$\boldsymbol{\gamma}(n) =  \nslash$, cf.\ \cite{DimockD}.) Note that $\dirop$
is no longer symmetric with respect to that scalar product. Up to sign (again depending
on metric signature), $<f,h>$ can be regained from $(f,h)$ as
$<f,h>  \ = (f,\boldsymbol{\gamma}(n)h)$. Thus, in setting up a framework for
Lorentzian spectral geometry, it is suggestive to add as another element
of structure an operator $\beta$ which, in the case of a Lorentzian spin
manifold, plays the role of $\boldsymbol{\gamma}(n)$ for some timelike, normalized
vector field $n$. This also induces the scalar product (\ref{scalprod}). When we
write $L^2$ space of spinors below, we are referring to this --- $n$-dependent ---
scalar product on the spinors. (Alternatively, one could work with the indefinite
inner product corresponding to $<f,h>$ for a Lorentzian manifold; this route
is taken in \cite{Strohmaier}.)

With these remarks in mind,
we present our proposal for the structure of Lorentzian spectral geometry.
We procced in such a way that we put side to side the objects forming
what we call a Lorentzian spectral triple (left column) and what they correspond
to in the ``classical case'', i.e.\ for a given Lorentzian spin manifold (right column).
\\[6pt]
A {\it Lorentzian Spectral Triple} (LOST) is a collection objects as follows:
$$ {( \mathcal{A}_0 \subset \mathcal{A}_2 \subset \mathcal{A}_b, \mathcal{H},D,\beta,
\overset{\circ}{\gamma},J)} $$
with the properties:
\begin{multicols}{2}
{\bf general (NC)}
\newline ${}$ \newline
${\mathcal{H}}$ is a Hilbert space
\par \columnbreak
{\bf classical}
\newline ${}$ \newline
${\mathcal{H} = L^2}$ spinors on a Lorentzian mani\-fold ${M}$ with spin structure
\end{multicols}
${}$ \dotfill
\begin{multicols}{2} \noindent
${\mathcal{A}_0}$ is a pre-${C^*}$-algebra \\
of bounded linear operators on ${\mathcal{H}}$,\\
${\mathcal{A}_b}$ is a preferred unitalization
${}$ \par
\columnbreak
${}$ \\
${}$ \  ${\mathcal{A}_0 = C_0^\infty(M)}$, ${\mathcal{A}_b = C^\infty_b(M)}$
\end{multicols}
${}$ \dotfill
\begin{multicols}{2} \noindent
$D$ is a linear operator with \\
dense ${C^\infty}$ domain
${\mathcal{H}^\infty = \mathcal{A}_2\mathcal{E}}$\\ with
a finitely generated 
\\ ${\mathcal{A}_b}$-module ${\mathcal{E}}$
\par
\columnbreak \noindent
$D = \dirop$ = Dirac-operator, with ${C^\infty}$ domain of
smooth sections in the spinor bundle where all $D$-derivatives are ${L^2}$
\end{multicols}
${}$ \dotfill
\begin{multicols}{2} \noindent 
${\beta , \overset{\circ}{\gamma}}$ are bounded operators \\
on ${\mathcal{H}}$, ${J}$ is anti-unitary,\\ with relations:\\
${ \beta^* = -\beta}$, 
${\beta^2= -1}$, \\
${ D^* = \beta D \beta}$,
${ [J a J,b]= 0 \ (a,b \in \mathcal{A}_0)}$, \\
$[X_1,[X_2,...[X_n,a]...]]$ are bounded for 
$a \in \mathcal{A}_0$ and $X_j = D$ or $D^*$\\ 
and several other relations
\par
\columnbreak \noindent
${ \beta} = \boldsymbol{\gamma}(n)$, where
${n}$ a timelike vectorfield on 
${M}$, \\
 $\boldsymbol{\gamma}(\,.\,)$ is the Clifford algebra action
on the spinor bundle,\\
$ \overset{\circ}{\gamma} = \boldsymbol{\gamma}(e_0) \cdots \boldsymbol{\gamma}(e_m)
$ with an orthonormal frame $(e_0,\ldots,e_m)$, \\
${J}$ corresponds to charge conjugation on the spinors\\
$D$ is a first order PDO
\end{multicols}
${}$ \dotfill
\begin{multicols}{2} \noindent
Setting $\langle D\rangle = \sqrt{D^*D + DD^*}$, \\
${a(1 - \langle D\rangle)^{-1}}$ is compact 
\\ for
$a \in \mathcal{A}_0$. 
\\
There is a minimal ${m \in \mathbb{N}}$
so that the Dixmier-trace of ${a \langle D \rangle^{-m}}$
is finite and non-vanishing for all ${a \in \mathcal{A}_0}$.
\par
\columnbreak \noindent
${}$ \\ ${}$ \\
${ \langle D \rangle}$ is
elliptic, $m$ is the (spectral) dimension of ${M}$.
\end{multicols}
${}$ \dotfill
\begin{multicols}{2} \noindent
plus 3 more conditions: $\overset{\circ}{\gamma}$ is
the image of a Hochschild cycle, ${\beta}$ belongs to the
1-forms of ${\mathcal{A}_b}$, and Poincar\'e duality
 (alternatively, closedness and Morita-equivalence of ${\mathcal{A}_b}$
 via ${\mathcal{E}}$)
\par
\columnbreak \noindent
${}$ \\
essentially: \\ orientability and Hodge-duality
\end{multicols}
We remark that this list of conditions on the objects of a LOST, as 
given here, is incomplete, and there are also some open questions,
e.g.\ related to precise domain conditions for $D$ and $D^*$
in relation to the inclusion of algebras 
$\mathcal{A}_0 \subset \mathcal{A}_2 \subset \mathcal{A}_b$.
In the
case of a Lorentzian spin manifold, one expects that one needs to impose regularity
conditions on the timelike vector field $n$ entering the
definition of $\beta = \boldsymbol{\gamma}(n)$ in order that good domain
conditions --- meaning that they lead to a reconstruction theorem of a
Lorentzian spin manifold in the case of an abelian $\mathcal{A}_b$ as
we will formulate it below --- can be obtained. 

As was emphasized, the structure of a LOST makes reference to 
a distinguished normalized, timelike vector field in the ``classical'' case.
However, the structure of a Lorentzian spin manifold does not single out
any preferred timelike vector field (except for special cases). In other words,
all LOSTs leading to the same (or isomorphic) Lorentzian spin manifolds
are to be viewed as equivalent. The following definition provides, in this sense,
the concept of equivalent LOSTs (where we use the symbol $\mathcal{A}$ as
abbreviation of the inclusions $\mathcal{A}_0 \subset \mathcal{A}_2 \subset \mathcal{A}_b$):
\\[10pt]
Let ${(\mathcal{A},\mathcal{H},D,\beta,\overset{\circ}{\gamma},J)}$ and
${ ( \widetilde{\mathcal{A}},\widetilde{\mathcal{H}},\widetilde{D},\widetilde{\beta},
\overset{\circ}{\gamma}\widetilde{\ },\widetilde{J})}$ be two LOSTs. 
\\[2pt]
They are called {\it equivalent} if there is a unitary ${U : \mathcal{H}
\to \widetilde{\mathcal{H}}}$ so that 
$$ { U {X} U^{-1} = \widetilde{{X}} \quad \text{for} \quad
{X} = \mathcal{A},\ \beta,\ \overset{\circ}{\gamma},\ J}$$
and
$$ { [\widetilde{D},U\,.\,U^{-1}] = U[D,\,.\,]U^{-1} }\,.$$
We expect that the open points in the definition of a LOST, which we alluded
to above, can be filled in such a way that the following conjecture can eventually be
established as a rigorous result.
\\[10pt]
{\it Conjecture} \\[2pt]
To each Lorentzian manifold with spin structure there corresponds
a LOST with commutative ${\mathcal{A}}$ and $D = \dirop$ = Dirac operator.
The LOSTs corresponding to isometric Lorentzian manifolds with equivalent spin structures
are equivalent.\\
Conversely, if a LOST has abelian ${\mathcal{A}}$, then it derives
from a Lorentzian manifold with spin structure. The Lorentzian spin manifolds deriving from
equivalent LOSTs are isometric and have equivalent spin structures.
\\[10pt]
The latter statement would amount to a Lorentzian version of Connes' reconstruction
theorem of Riemannian spin manifolds for abelian $\mathcal{A}$. The idea is, of 
course, to use the structure of a LOST to derive from it a Riemannian spectral
triple from which the manifold structure can be constructed as in \cite{Connesbigpaper}.
However, if this can be achieved is, as yet, still open.
\section{GHYSTs and Quantum Field Theory}
The LOST setting has the potential to describe very general Lorentzian spin
manifolds and their NC generalizations. According to our present understanding,
however, it is most difficult to set up a consistent framework for 
quantum field theory on
Lorentzian
spin manifolds which are not globally hyperbolic \cite{HawkingCPC,Kay,KRW}. Therefore, 
in order to achieve a promising framework for quantum field theory on LOSTs,
a first step consists in characterizing the counterpart of global hyperbolicity
at the LOST level.

A Lorentzian spin manifold $M$ is globally hyperbolic if and only if the Dirac operator
$\dirop$ defined on it possesses unique advanced and retarded fundamental
solutions, $R_{+}$ and $R_{-}$, taking
$C_0^\infty$ sections in the spinor bundle to $C^\infty$ sections. They are characterized by
\begin{align} \label{fund}
 R_{\pm} \dirop f  & = f = \dirop R_{\pm}f 
\end{align} 
for all $C_0^\infty$ spinor sections $f$, and by 
\begin{align} \label{supp}
{\rm supp} (R_{\pm} f) & \subset J^{\pm}({\rm supp}(f))\,,
\end{align}
where $J^{\pm}(S)$ denotes the causal future $(+)$ / causal past $(-)$ set
of a subset $S$ of the Lorentzian spin manifold. This means, $J^{\pm}(S)$
is the set of points in $M$ which can be reached by all future $(+)$ / past $(-)$
directed causal curves emanating from $S$. 

Therefore, to characterize globally hyperbolic LOSTs --- which will be referred to
as GHYSTs, short for {\it globally hyperbolic spectral triples} --- one would
have to formulate conditions characterizing advanced and retarded fundamental
solutions for the operator $D$ in the LOST setting, i.e.\ using only the objects
forming a LOST. Clearly, condition (\ref{fund}) can readily be generalized to
the LOST setting. But condition (\ref{supp}) uses the localization concept
of a ``classical'' differentiable manifold and this is not at hand within the LOST
setting. Hence, it is unclear how condition (\ref{supp}) should be generalized
to the LOST setting, and how to characterize advanced and retarded fundamental
solutions of $D$ in this setting. Nevertheless, let us for the moment proceed
under the hypotheses that a suitable characterization of advanced and retarded
fundamental solutions of $D$ in the LOST setting can be given. At the level of
concrete examples, there are situations where there are
obvious candidates for advanced and retarded fundamental solutions:
 Drawing largely on results of \cite{GGBISV}, it can be shown that Moyal-Minkowski
spacetime 
(a description of whose basic elements will be given in the next section)
is an example of a LOST (M.\ Paschke, unpublished), and in that 
particular case,
$D$ is just the usual Dirac operator on Minkowski spacetime which has unique
advanced and retarded fundamental solutions in the ``classical'' sense. 
In fact, as a consistency requirement, the concept of adavanced and retarded
fundamental solutions of $D$ in the LOST setting should coincide with the ``classical''
concept whenever a LOST describes a Lorentzian spin manifold.  
In the following, we shall take it for granted
(more appropriately, take as a working hypothesis) that Moyal-Minkowski spacetime
is a GHYST.  

Now suppose we have a GHYST 
$$ {\bf G} = {(\mathcal{A},\mathcal{H},D,\beta,\overset{\circ}{\gamma},J,R_{\pm})}$$
where $R_{\pm}$ are the advanced and retarded fundamental solutions of $D$. Furthermore,
setting $R = R_+ - R_-$
suppose --- as is the case for a globally hyperbolic spin manifold --- that
$$ (f,h)_{(R)} = (f,\beta R h)  $$
for $f,h$ in a suitable dense domain  $\mathcal{H}_{(R)}$ 
in $\mathcal{H}$ is positive semi-definite (possibly
up to a constant overall phase). We denote by $\mathcal{K}$ the
completion of $\mathcal{H}_{(R)}$ factorized by the kernel of
$(\,.\,,\,.\,)_{(R)}$ with respect to the scalar product
$(\,.\,,\,.\,)_{\mathcal{K}}$ induced
by $(\,.\,,\,.\,)_{(R)}$. One can show that $J$ furnishes a conjugation
on $\mathcal{K}$ (again denoted by $J$). Moreover, 
$R: f \mapsto Rf$ is, under suitable identification, equivalent to the  
canonical surjection $\mathcal{H}_{(R)} \to \mathcal{K}$.
Thus, one can invoke the abstract CAR quantization procedure \cite{Araki}
to associate to ${\bf G}$ an abstract $C^*$-``Dirac field''-algebra ${\sf F}({\bf G})$
which is generated by a family $B(\chi)$, $\chi \in \mathcal{K}$, subject
to the conditions
\begin{itemize}
\item $ \chi \mapsto B(\chi)$ is $\mathbb{C}$-linear
\item $B(\chi)^* = B(J\chi)$
\item $ B(\chi)^*B(\xi) + B(\xi)B(\chi)^* = (\chi,\xi)_{\mathcal{K}} {\bf 1}$
\end{itemize}
where ${\bf 1}$ denotes the unit element in the $C^*$-algebra ${\sf F}({\bf G})$.
Upon setting $\Psi(f) = B(R f)$ for $f \in \mathcal{H}_{(R)}$ (identifying
$R$ with the canonical surjection), one obtains an ``abstract Dirac field''
over ${\bf G}$ with the characteristic conditions
\begin{align*}
\Psi(f)^*\Psi(h) + \Psi(h)\Psi(f)^* & = (f,h)_{(R)} {\bf 1} \quad (f,h \in \mathcal{H}_{(R)})\quad
 \text{and} \\
\Psi(Df) & = 0 \,.
\end{align*}
The latter equation corresponds to an ``abstract Dirac equation'' associated to
the underlying GHYST ${\bf G}$. 

Finally, the assignment ${\bf G} \to {\sf F}(\bf G)$ of GHYSTs to $C^*$-CAR algebras
is functorial in the following sense.
Let us call a unitary equivalence
$$ {\bf G} \overset{U}{\longrightarrow} \widetilde{\bf G} $$
{\it rigid} if $U D U^* = \widetilde{D}$. Then
there is a canonical $C^*$ algebraic morphism 
$$ {\sf F}({\bf G}) \overset{\alpha_U}{\longrightarrow} {\sf F}(\widetilde{\bf G}) $$
which is induced by $\alpha_{U}(\Psi(f)) = \widetilde{\Psi}(U f)$ in obvious
notation. This implies the covariance property $\alpha_{U_2} \circ \alpha_{U_1}
= \alpha_{U_2 U_1}$ for rigid unitary equivalences.

This functorial structure corresponds to the ``global'' covariance of the
quantized Dirac field on globally hyperbolic spacetimes which was first brought to the
fore by J.\ Dimock \cite{DimockD}. The Dirac field fulfills also a stronger, ``local'' version
of covariance
\cite{VerSPST,Sanders,BFV,BaerGinoux} which induces essentially the local
and causal structure of the quantum field theory. However, this ``local covariance'',
which to a large part also determines the interpretation of the  quantum
field theory (derived from the net of local observable
algebras cf.\ \cite{Haag}), crucially depends on the localization concept
of classical differentiable manifolds, and that has no direct counterpart in the framework
of LOSTs or GHYSTs. How, then, does one link the non-commutativity
of the $\mathcal{A}$ algebra of ${\bf G}$ with the algebraic structure of
${\sf F}({\bf G})$, and which quantum field operators
associated to ${\bf G}$ carry a 
particular physical interpretation? While one can surely come up with
many ideas for answers, we actually take a modest step and
look at the simplest example of the ``abstract'' quantized Dirac
field on a GHYST with NC $\mathcal{A}$ --- corresponding to Moyal-Minkowski
spacetime.

\section{Dirac field on Moyal-Minkowski spacetime}
Moyal-Minkowski spacetime is identical to Minkowski spacetime, except that
the usual commutative pointwise product of (Schwarz-class) test-functions
on Minkowski spacetime is replaced by the Moyal-product (or Rieffel-product).
To set up matters more precisely, consider $n =1+d$ dimensional Minkowski
spacetime $\mathbb{R}^{1,d}$. Let $\Theta = (\Theta_{\mu \nu})$ be an anti-symmetric
real $n \times n$ matrix. Then one can define a deformed product of 
$\mathcal{S}(\mathbb{R}^n)$ by
\begin{align} \label{Mpro}
 f \star h (x) = (2\pi)^{-n} \int \int f(x -\frac{1}{2}\Theta u)h(x + v)
 {\rm e}^{-i u \cdot v}\,d^n u\,d^n v 
\end{align} 
where $u \cdot v$ is the standard euclidean scalar product on $\mathbb{R}^n$.
Usually, when $n$ is even, one takes $\Theta$ to be the standard symplectic
matrix times a positive scaling factor. In this case, and also in more general cases, one
can show that the above product between test-function is associative.
When $\Theta$ has non-zero entries it is, however, non-commutative.

We now define an algebra $\mathcal{A}_0 = \mathcal{S}(\mathbb{R}^n)$ with
the above Moyal product as algebra product. By the $\star$-product it acts naturally
on the Hilbert-space $\mathcal{H} = L^2(\mathbb{R}^n,\mathbb{C}^{N})$
where $N = N(n)$ is the lowest dimension for an irreducible, self-dual representation of the
Clifford algebra $Cl(1,d)$ (requiring existence of such a representation puts
restrictions on the dimension $n$, see \cite{BorVe} and references
given there for details).  The Clifford
algebra generators are then represented by a set of ``Gamma-matrices'',
$\gamma^0,\gamma^1,\ldots,\gamma^d$. When taking $D$ as the
usual Dirac-operator (with some arbitrary, but fixed mass term $m < 0$),
$$ D = i\gamma^\mu \frac{\partial}{\partial x^\mu} + m $$ 
on the domain $\mathcal{S}(\mathbb{R}^n,\mathbb{C}^N)$, together
with $\beta = \gamma_0$, $\overset{\circ}{\gamma} = \gamma_0 \cdots \gamma_d$,
$J$ as the charge-conjugation on $\mathcal{H}$, and taking $\mathcal{A}_2$
and $\mathcal{A}_b$ as in \cite{GGBISV}, then one can use much of the
results of \cite{GGBISV} to show that one has collected the data of a LOST,
at least in the case of even $n$ and with non-degenerate $\Theta$
(M.\ Paschke, unpublished --- as there is no complete published proof,
we now proceed under the fiction that these data indeed form the data
of a LOST). This LOST is essentially just the LOST corresponding to
$n$-dimensional ``classical'' Minkowski spacetime, but with the Moyal product
instead of the commutative, pointwise product on the algebra of test-functions. It is even
a GHYST (strictly, we assume the conditions on GHYSTs to be formulated
so that this holds true, cf.\ our discussion above), since the Dirac
operator $D$ possesses unique advanced and retarded fundamental solutions
(in the classical manifold sense, not (yet) expressed using only the data of
the LOST). Using these advanced and retarded fundamental solutions
$R_\pm$ and their difference $R = R_+ - R_-$, one can even set up the
CAR algebra ${\sf F}_{\rm MM}$ of 
the quantized Dirac field on
Moyal-Minkowski spacetime. However,
it is easy to see that this algebra is in no way different from the 
CAR algebra ${\sf F}_{\rm Mink}$ of the quantized Dirac field. The
information about the non-commutativity of $\mathcal{A}_0$ is
not directly visible in at the level of the generators of the algebras --- i.e.\ in
the quantization procedure, if one wishes to put it like that --- but it 
is hidden somewhere else. How can we access this information? Which
Dirac quantum field operators carry that information? Does it provide
any link to the more customary approach of quantum field theory on
Moyal deformed spaces which essentially replaces the ``usual'' products of
Dirac quantum field operators by their Moyal-Rieffel products? 

 Obviously, we need to look at some way the elements $c$ of the algebra
of test-functions can take action on the Dirac quantum field operators.
Let $\Psi(f)$ denote the abstract quantum field operators ($f \in \mathcal{S}
(\mathbb{R}^n,\mathbb{C}^N)$) generating both ${\sf F}_{\rm MM}$ and
${\sf F}_{\rm Mink}$. On usual Minkowski spacetime, one can look at
the map $\Psi(f) \mapsto \Psi(c Rf)$. This map arises in when scattering
the quantized Dirac field by an external scalar potential $c$. In the
next sections, we explain this, and explain how this potential scattering
can be generalized to scattering by an NC potential.

\section{Dirac field NC potential scattering, \\
1: Commutative time}

In order to keep matters as simple as possible, we will, in the present
Section, specialize to the case $n = 3$ (implying $N=4$), i.e.\ 3-di\-men\-sio\-nal Minkowski-
resp.\ Moyal-Minkowski spacetime. However, most of our considerations
apply to more general spacetime dimensions, see \cite{BorVe} for details.

To begin, we need a bit of notation. We denote by $\mathcal{K}_{+}$ the
positive frquency part of the
one-particle Hilbert space of the Dirac field on 3-dimensional Minkowski
spacetime. This corresponds to the subspace of 
positive frequency solutions of 
the solution space
$\mathcal{K}$ (containing solutions $\chi$, with Cauchy-data of Schwarz class, to the 
Dirac equation $(i \gamma^\mu \partial_{x^\mu} + m)\chi = 0$)
\cite{Thaller}.
Furthermore, we consider the quantized Dirac field on Minkowski spacetime
in its usual vacuum representation. Consequently, we identify the ``abstract
Dirac field operators'' $\Psi(f)$ with the represented operators
$\psi(f)$ which are concretely given, as usual, in terms of annihilation
and creation operators in the fermionic Fock space $F_+(\mathcal{K}_{+})$
over the one-particle space $\mathcal{K}_{+}$. 

Now let $c$ be a real-valued Schwarz function on $\mathbb{R}^3$. Then
one can show \cite{BorVe} --- and we believe it is well-known --- that
\begin{align} \label{commu}
i[:\psi^+\psi:(c), \psi(f) ]  &=  \psi(cRf) 
\end{align}
holds for all test spinors $f \in \mathcal{S}(\mathbb{R}^3,\mathbb{C}^4)$.
Here, 
$[X,Y] = XY -YX$ denotes the commutator, and
$:\psi^+\psi:(c)$ is the normal-ordered coinciding-point-limit-product
(Wick-product) of the Dirac-adjoint $\psi^+$ with $\psi$ itself. This
results in a scalar quantum field, which in the above formula is smeared
with $c$ as a test function (see \cite{BorVe} for further discussion).

On the other hand, (\ref{commu}) is the result of differentiating the scattering
transformation related to external potential scattering of the quantized Dirac field
on Minkowski spacetime with respect to the potential strength.
Let us explain this in a bit more detail. 

First, we put the free Dirac equation $(i \gamma^\mu \partial_{x^\mu} + m)\chi =0$
into Hamiltonian form: We fix some inertial time coordinate $t$ ( $\equiv x^0$)
and write $\chi_t(\,.\,) = \chi(t,\,.\,)$.  Then the free Dirac equation is equivalent to
$$ i \frac{d}{dt} \chi_t + H_0\chi_t = 0 $$
where the free Hamiltonian is a selfadjoint operator on a suitable dense domain
in $L^2(\mathbb{R}^2,\mathbb{C}^4)$ which acts as
$$ H_0 v(\underline{x}) = (i \gamma^0\gamma^k \partial_{x^k} + \gamma^0 m) v(\underline{x}) $$
where $\underline{x} = (x^k)_{k=1}^2$. 
Now let $c = c(t,\underline{x})$ be a real-valued Schwarz function, regarded as 
a time-dependent external scalar potential for the Dirac field. Then
the Dirac equation 
$$ (D + \lambda c)\chi =  (i \gamma^\mu \partial_{x^\mu} + m + \lambda c)\chi = 0$$
is equivalent to 
\begin{align} \label{tdhe}
H_\lambda (t)\chi_t & = (H_0 + V_{\lambda}(t))\chi_t = 0
\end{align}
with the time-dependent potential operator
\begin{align} \label{tdpot}
 V_\lambda (t)v(\underline{x}) = \lambda\gamma^0 c(t,\underline{x})v(\underline{x})
\end{align}
defined on a suitable domain of $L^2(\mathbb{R}^2,\mathbb{C}^4)$; here
we have introduced a positive real parameter $\lambda$ scaling the
strength of the interaction with the external potential.
Assuming that appropriate self-adjointness and domain conditions are
fulfilled (see \cite{BorVe} and literature cited there for full details), one can
show that there is a two-parametric family of unitaries $U_\lambda (t,s)$ in
$L^2(\mathbb{R}^2,\mathbb{C}^4)$ with
$U_\lambda (t,r)U_\lambda (r,s) = U_\lambda (t,s)$, $U_\lambda (t,t) = {\bf 1}$ and such that 
$\chi_t = U_\lambda (t,t_0)v$ is the unique solution to (\ref{tdhe}) with initial
condition $\chi_{t_0}(\underline{x}) = v(\underline{x})$.  Moreover,
the one-particle scattering operator 
$$ s_{\lambda} = \lim_{\pm t_{\pm} \to \infty} \,
{\rm e}^{it_+H_0}U_\lambda (t_+,t_-){\rm e}^{-it_-H_0}$$
exists and is a unitary on the space $L^2(\mathbb{R}^2,\mathbb{C}^4)$
of Cauchy-data for the free Dirac equation. The latter Hilbert space is
canonically isomorphic to the solutions' Hilbert space $\mathcal{K}$. 
Remembering that $\psi(f)$ depends only on $Rf$, where $R$ is the
difference of advanced minus retarded fundamental solution of the 
free Dirac equation, one can define a re-labelled field operator $\check{\psi}(Rf) = \psi(f)$,
and identifying $s_\lambda$ with an operator in $\mathcal{K}$, one
can actually show that 
$$ \frac{1}{i} \left. \frac{d}{d\lambda}\right|_{\lambda = 0} {\psi}(s_\lambda Rf)
 = \check{\psi}(cRf) \,.$$
Hence, using (\ref{commu}), one finds
$$ [:\psi^+\psi:(c), \check{\psi}(Rf)] = 
\frac{1}{i} \left. \frac{d}{d\lambda}\right|_{\lambda = 0} \check{\psi}(s_\lambda Rf) \,.$$
Moreover, the one-particle scattering transformation $s_\lambda$ can be 
unitarily implemented in the vacuum representation of the free Dirac field, meaning
that there is a unitary $S_\lambda$ on $F_+(\mathcal{K_{0+}})$ (the S-matrix,
or 2nd quantized scattering operator) such that
$$ S_\lambda \check{\psi}(Rf) S_\lambda^* = \check{\psi}(s_\lambda Rf)\,.$$
Therefore, one has
$$  [:\psi^+\psi:(c), \check{\psi}(Rf)] = 
\frac{1}{i} \left. \frac{d}{d\lambda}\right|_{\lambda = 0} S_\lambda \check{\psi}(Rf) S_\lambda^* \,.$$

Now we will see that one obtains identical results when replacing the potential operator
$V_\lambda(t)$ of (\ref{tdpot}) by a more general operator involving the Moyal product,
at least as long as our underlying 3-dimensional Moyal-Minkowski spacetime still
has ``commutative time''. (This restriction will be lifted in the next section.)
As the matrix appearing in the definition of the Moyal product we choose
$$ \Theta =(\Theta^{\mu\nu}) = \left( \begin{array}{ccc}
                   0 & 0 & 0 \\
                   0 & 0 & \theta \\
                   0 & -\theta & 0 \end{array} \right) $$
where $\theta$ is some positive parameter that will be kept fixed.                   
Let us now choose some real-valued scalar test function $c$ of the form
$$ c(t,\underline{x}) = a(t)b(\underline{x}) $$
where $a$ is $C_0^\infty$ and $b$ is Schwarz. Then we define the two 
interaction potentials
\begin{align}
 V_{\lambda\, \star}(t) v(\underline{x}) & = \lambda a(t)\gamma^0( b \,\underline{\star}\, v (\underline{x})
  + v \,\underline{\star}\, b (\underline{x})) \,, \\
  V_{\lambda\, \star \star}(t) v (\underline{x})
   & = \lambda a(t)^2 \gamma^0 ( b \,\underline{\star}\, v\,\underline{\star}\, b (\underline{x}) ) 
\end{align}
where $\underline{\star}$ denotes the Moyal product on Schwarz functions over
$\mathbb{R}^2$, given by
\begin{align}
   b\, \underline{\star}\, g (\underline{x}) & = (2\pi)^{-2} \int \int b(\underline{x} -\frac{1}{2}
\underline{\Theta} \underline{y})g(\underline{x} + \underline{q})
 {\rm e}^{-i \underline{y} \cdot \underline{q}}\,d^2 \underline{y}\,d^2 \underline{q} 
\end{align}
with
$$ \underline{\Theta} = \left( \begin{array}{cc}
                                     0 & \theta \\
                                     -\theta & 0
                                     \end{array} \right) \,. $$
Thus, if $\chi_t(\,.\,) = \chi(t,\,.\,)$ is a solution to
$$ i \frac{d}{dt} \chi_t + V_{\lambda\, \#}(t)\chi_t =0\ \quad (\# = \star \ \text{or}\ \star \star)\,,
$$
this is equivalent to 
\begin{align}
 D\chi  + \lambda( c \star \chi + \chi \star c ) &= 0 \quad \text{if} \ \# = \star\,,\ \text{and} \\
 D\chi + \lambda c \star \chi \star c & = 0 \quad \text{if} \ \# = \star \star\,.
\end{align}
In \cite{BorVe}, we have established the following results.
\\[10pt]
\begin{Theorem} ${}$ \\[2pt]
\begin{itemize}
\item The one-particle scattering operator $s_{\lambda\,\#}$ exists for the
potentials $V_{\lambda\, \#}(t)$ defined above ($\# = \star$ or $\# = \star\star$).
\item The one-particle scattering operator is unitarily implemented in the
vacuum representation of the quantized Dirac field, i.e.\ there are unitary
operators $S_{\lambda\,\#}$ on the Fock-space $F_+(\mathcal{K}_{0+})$ such
that 
$$ S_{\lambda\, \#} \check{\psi}(Rf) S_{\lambda\, \#}^*
 = \check{\psi}(s_{\lambda\,\#} Rf) \,.$$
\item There is an essentially self-adjoint operator 
$\Phi_{\#}(c)$ on the Wightman domain of Fock-space such that
\begin{align}
i[\Phi_{\#}(c) , \psi(f)] & =  \left. \frac{d}{d\lambda} \right|_{\lambda = 0}
                           S_{\lambda\, \#} \psi(f) S_{\lambda\, \#}^* \\
 & =
 \begin{cases}
  \psi (  c \star Rf + Rf \star c) \quad \text{if} \ \# = \star \\
  \psi( c \star Rf \star c) \quad \text{if} \ \# = \star \star
 \end{cases} 
\end{align}
\end{itemize}
\end{Theorem} 

\section{Dirac field NC potential scattering, \\
2: The general case}

In the present section, our aim is to generalize the findings on the
scattering of the quantized Dirac field by an NC potential, but still keeping
time ``commutative'', to the general case, where also time is turned
into an NC ``coordinate (operator)''. To this end, we now consider 
4-dimensional Moyal-Minkowski spacetime
(but the discussion of this section can
be generalized to other spacetime dimanesions,
see \cite{BorVe2}. The Moyal-product of test-functions
$f$ and $h$ on $\mathbb{R}^4$ is as in (\ref{Mpro}) for $n=4$, with the
matrix 
$$ 
 \Theta = \theta \left(
\begin{array}{cccc} 
  0 & 1 & 0 & 0\\
  -1 & 0 & 0 & 0\\
  0 & 0 & 0 & 1\\
  0& 0 & -1 & 0 \end{array} \right)
 $$
where $\theta$ is some fixed positive parameter. 

For simplicity, we consider only one of the non-commutative potential
terms from the two of the previous section --- corresponding to the
field equation
\begin{align} \label{DEncT}
D\chi + \lambda c \star \chi \star c & = 0
\end{align}
for some real-valued Schwarz function on $c$ on $\mathbb{R}^4$. (Again,
$\lambda$ is a positive parameter scaling the interaction coupling.)
Now we face the problem that, due to the non-local action of the Moyal
product with respect to the time-coordinate, the field equation is no
longer equivalent to a time-dependent Hamiltonian equation 
where at each point in time the Hamilton operator acts
only with respect to the spatial coordinates. Thus, we need another
way of finding solutions to (\ref{DEncT}). The first step is to not consider
(\ref{DEncT}) as it stands, but to replace it by a simpler form where
the potential is made nicer by introducing suitable cut-offs, then to 
establish solutions to the cut-off dynamical equations, and finally to control the
limit of such solutions as the cut-offs are being removed.

In fact, we consider two cut-offs. Let $\tau > 0$ and let, with respect to
a chosen time-coordinate $t$, $M_{\tau} = \{ (t,x^1,x^2,x^3) \in \mathbb{R}^4: 
 -\tau < t < \tau\}$ be a slice of Minkowski-spacetime whose time-extension
 is controlled by $\tau$. 
Then consider any non-negative $C_0^\infty$ function $\xi$ defined on $\mathbb{R}$
which is equal to 1 on the interval $[-\tau/2,\tau/2]$ and zero outside the
interval $(-\tau/\sqrt{2},\tau/\sqrt{2})$. Then we define the operator
$D + V_\xi(\lambda)$ on the spinor fields $\mathcal{S}(M_\tau,\mathbb{C}^4)$
over $M_\tau$
(defined as having compact support in time) where the cut-off potential operator
is given by
$$ V_\xi(\lambda)f = \lambda \xi( c \star (\xi f) \star c) \,.$$
(Here, $\xi$ acts as multiplication operator.) Then one can show that, provided
$\lambda$ is sufficiently small (depending on $\tau$ and $\xi$), there exist
unique advanced and retarded fundamental solutions $\mathcal{R}^\pm_{\tau,\lambda,\xi}$
for the operator $D + V_\xi(\lambda)$ on the slice $M_\tau$ which can be
gained as Neumann series \cite{BorVe2},
$$  \mathcal{R}^{\pm}_{\tau,\lambda,\xi} = 
R^\pm (1 + \lambda V_\xi(c)R^\pm )^{-1} = R^\pm \left(\sum_{j = 0}^\infty
 (-1)^j (\lambda V_\xi(c)R^\pm)^j\right) $$
where $R^\pm$ denote the advanced/retarded fundamental solutions of
the Dirac operator $D$ on $M_\tau$. Using these advanced/retarded fundamental
solutions, it is possible to define a one-particle scattering operator
$s = s(\tau,\lambda,\xi)$ on the solution space of the free Dirac equation
on $M_\tau$. Schematically, the action
of this one-particle scattering operator can be described
as follows. One chooses initial data for the free Dirac
equation at $t=0$, propagates those data forward in time with the
dynamics of the free Dirac equation up to $t= \tau/1.25$; then, using 
$\mathcal{R}^{\pm}_{\tau,\lambda,\xi}$, one propagates the initial data
backwards in time using the dynamics of $D + V_\xi(\lambda)$ up to
$t = -\tau/1.25$, and then the resulting initial data are propagated forward
in time back to $t=0$, using the dynamics of the free Dirac equation.
What we just described verbally is depicted in the diagram of Figure 1,
when following the arrows counterclockwise starting from the solid black line
which represents the $t=0$ hyperplane in $M_\tau$. 
\begin{center}
\includegraphics[width=12cm]{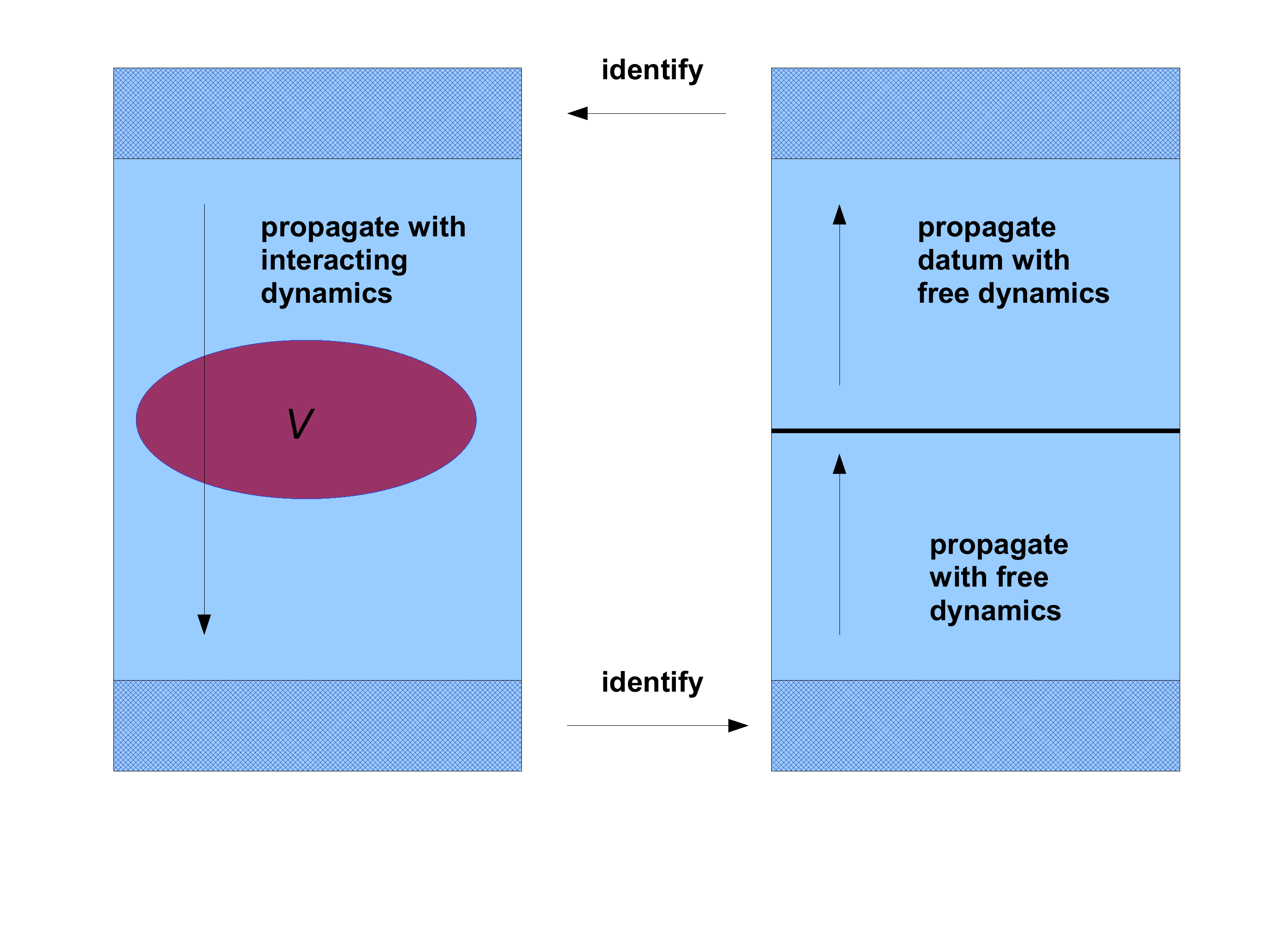}
\end{center}
{\small
\begin{center}
{\bf Figure 1. Sketch of action of the one-particle scattering operator
for the cut-off dynamics.}
\end{center}
 }
By standard arguments \cite{Araki,BR2}, the one-particle scattering operator
 $s(\tau,\lambda,\xi)$ induces a $C^*$-algebraic Bogoliubov-transformation
$\beta_{\tau,\lambda,\xi}$ on ${\sf F}(M_\tau)$, the CAR-algebra of
the free Dirac field on $M_\tau$ (regarded as a spacetime in its own right),
by setting
$$ \beta_{\tau,\lambda,\xi}(\Psi(f)) = \check{\Psi}(s(\tau,\lambda,\xi) Rf) \,.$$
Differentiating with respect to $\lambda$, one obtains a derivation $\delta_{\tau,\xi}$
on ${\sf F}(M_\tau)$,
\begin{align*}
\delta_{\tau,\xi}(\Psi(f)) & =
 \left. \frac{d}{d\lambda} \right|_{\lambda = 0} \beta_{\tau,\lambda,\xi}(\Psi(f)) \\
  & = \Psi (\xi( c \star ( \xi Rf) \star c)) \,.
\end{align*}
Finally, one can remove the cut-offs by letting $\tau \to \infty$ and $\xi \to 1$.
The following result states that the these limits are well-behaved.
\begin{Proposition}
(a) The limit of $\delta_{\tau,\xi}(\Psi(f))$ as $\tau \to \infty$, $\xi \to 1$ exists
for each test-spinor $f \in \mathcal{S}(\mathbb{R}^4,\mathbb{C}^4)$. It defines
a derivation $\delta$ on ${\sf F}(\mathbb{R}^4)$, the CAR-algebra of the
free Dirac field on Minkowski spacetime, acting as 
$$ \delta(\Psi(f)) = \Psi( c \star Rf \star c) \,.$$
\\[4pt]
(b) There is an essentially selfadjoint operator $Y(c)$ on the Wightman domain
of the Fockspace $F_+(\mathcal{K_{0+}})$ such that
$$ i[Y(c),\psi(f)] = \delta(\psi(f)) = \psi( c \star Rf \star c) $$
where $\psi(f)$ is the vacuum representation of $\Psi(f)$ on the Fockspace
$F_+(\mathcal{K_{0+}})$ in terms of creation and annihilation operators.
\end{Proposition}
The proof of this proposition will appear in \cite{BorVe2}.

\section{An Answer to a Previous Question (and some discussion)}

At the end of Section 4 we posed the question how one can retrieve  information
about the non-commutativity of the underlying Moyal-Minkowski spacetime
from the quantized Dirac field constructed on it by abstract CAR quantization
(of the GHYST describing Moyal-Minkowski spacetime), as apparently the construction
of the quantized Dirac field is no other than in the case of usual Minkowski spacetime.

In  Section 5 we have seen that for the case of usual Minkowski-spacetime,
for a scalar test-function $c$ the passage from $\psi(f)$ to $\psi(c Rf)$
is given by a derivation on the CAR algebra of the (free) Dirac field in vacuum representation,
$$ \psi( c Rf) = [ : \psi^+\psi:(c),\psi(f)]\,. $$
To be noted is, first, that the derivation is induced by a selfadjoint operator
$:\psi^+\psi:(c)$, the quantized counterpart of the absolute square of field strength
of the Dirac field, smeared with $c$. This operator thus gives a measure of
the  localization and strength of the external field inducing the 
scattering process. Secondly, the derivation is obtained by differentiating the
S-matrix with respect to the field strength scaling parameter $\lambda$, and 
following the idea of Bogoliubov's formula \cite{BogShi}, diffentiating an
S-matrix of an interaction with respect to the interaction coupling strength
is a general method of obtaining the observable quantum fields of a 
quantum field theory. 

For usual Minkowski spacetime, we view the test function $c$ as an element of
the commutative algebra $\mathcal{A}_0$ entering the data of the GHYST
corresponding to Minkowski spacetime, and therefore, we view $Rf \mapsto
c Rf$ as the algebraic action of that algebra on a suitable module. This
point of view we carried over, in Section 5, to the case of Moyal-Minkowski space
(with commutative time): Here, the algebra $\mathcal{A}_0$ are the test-functions
with the non-commutative Moyal product. The ``module actions'' are, therefore, modified
to $Rf \mapsto c \star Rf + Rf \star c$ or $Rf \mapsto c \star Rf \star c$. (Strictly speaking,
these aren't module actions; the symmetrized form here is needed to ensure
$J$-invariance of the resulting potential term in order to be able to obtain Bogoliubov
transformations on the CAR algebra.) In the case of commutative time studied in Section 5
we could use a variation of the methods used to solve the scattering problem
for a usual scalar potential to obtain a solution to the scattering problem
for the NC scalar potential. This also leads to an S-matrix, and differentiating with respect
to the field strength gives derivations induced by operators $\Phi_\# (c)$ such that
\begin{align*}
 i[\Phi_\star (c),\psi(f)] & = \psi( c \star Rf + Rf \star c) \,,\\
 i[\Phi_{\star \star} (c), \psi(f)] & = \psi( c \star Rf \star c) \,.
\end{align*}
It is important here that the $c$ appearing in the argument of $\Phi_\# (c)$ is to
be viewed not just as a test-function, but as an element of the non-commutative
algebra $\mathcal{A}_0$ of test-functions endowed with the Moyal product. Again,
$\Phi_\# (c)$ is an observable measuring strength and localization of the external
 --- and now, non-commutative --- potential, where the localization is, due to the
 non-local action of the Moyal product, no longer as sharp as in the sense
 of localization on a usual differentiable manifold.
Furthermore, the penultimate equation furnishes a link to the more heuristic approach
to quantum field theory on Moyal-Minkowski spacetime where the usual product
$AB$ of quantum field operators is replaced by the Rieffel-Moyal product
\cite{Rieffel,BLS},
$$ A \star_\Theta B =
  (2\pi)^{-n} \int \int \alpha_{\frac{1}{2}\Theta u}(A)\alpha_{-v}(B)
 {\rm e}^{-i u \cdot v}\,d^n u\,d^n v 
$$
where $\alpha_{\cdot }$ denotes the automorphic action $\alpha_y(\psi(f)) = \psi(f_y)$, $f_y(x) = f(x-y)$
of the translations on the operator algebra generated by the Dirac field in $n$-dimensional
Minkowski spacetime. For the sake of simplicity, let us assume that the parameter $\theta$
appearing in the definition of $\Theta$ is equal to 2. Then we have, formally,
$$ \psi(c \star Rf + Rf \star c) = i[:\psi^+\psi:(c), \psi(f)]_{\Theta} +
                                               i [:\psi^+\psi:(c), \psi(f)]_{-\Theta} \,.$$
The notation means that in the first commutator, the operator product is 
replaced by the product $\star_\Theta$, while in the second commutator,
the operator product is replaced by the product $\star_{-\Theta}$. As the equation
stands, it is only formal in nature because one cannot rely on the theorems
in \cite{Rieffel,BLS} for the existence of the product $\star_\Theta$ owing to the
fact that $:\psi^+\psi:(c)$ is an unbounded operator and $\Theta$ is degenerate in
the case of commutative time, so one would have to specify very carefully the domain
on which the equality is valid. Relegating this technical question elsewhere, one
can see that the Rieffel-Moyal product between quantum field operators appears
naturally in the present setting, too.

The results of Section 6 show that the vantage point just described can also be maintained
in the case  of non-commutative time. The central difficulty here is to define a dynamics
for the interaction potential which now is non-commutative and hence, non-local in
time, so that it cannot be formulated as a time-dependent Hamiltonian dynamics
as in Section 5. Nevertheless, interpreting the dynamical problem in terms of
a family of cut-off dynamics, one again obtains an operator $Y(c)$ which basically
can be seen as the result of differentiating the S-matrix with respect to the interaction
coupling strength of the non-commutative potential. Thus, it appears that in the
case of Moyal-Minkowski spacetime (as a --- strictly speaking, hypothetical --- model
for a GHYST) the CAR-quantization together with external scattering by a non-commutative
potential and ``Bogoliubov's formula'' yields a correspondence between 
(hermitean) elements $c$ in the algebra $\mathcal{A}_0$ and observables
$\Phi_\#(c)$ or $Y(c)$ of the quantized Dirac field, also establishing a relation
to the Rieffel-Moyal product between quantum field operators. Despite the fact
that in the construction of the S-matrix, or the related derivations
with respect to the potential coupling strength, we have used some properties which are not
direct consequences of the structure of a GHYST, like commutative time in 
Section 5 or the cut-off dynamics localized in time in Section 6, we are confident
that our construction of a relation between elements of $\mathcal{A}_0$ and
observable quantum field operators can in principle be extended to more general GHYSTs.
This of course requires a better understanding of the structure of GHYSTs, and in 
particular, of concepts of localization in non-commutative geometry.

\end{document}